%% file: main.tex
\documentclass[10pt, a4paper, twocolumn, showabstract]{naverlabseurope}

\usepackage{lipsum}
\usepackage{multicol}
\usepackage{tikz,tkz-kiviat,pgfplots}

\usepackage[inline]{enumitem}
\usepackage{pbox}
\usepackage{balance}
\newlist{inlinelist}{enumerate*}{1}
\setlist*[inlinelist,1]{label=\roman*),itemjoin={{, }},itemjoin*={{, and }}}
\usepackage{url}
\usepackage{multirow}
\usepackage{multicol}
\usepackage{booktabs}
\usepackage{tikz}
\usepackage{xcolor}
\usepackage{lipsum}
\usepackage{acronym}
\usepackage{subcaption}
\usepackage{xspace}
\usepackage{arydshln}
\usepackage{array}
\usepackage{adjustbox}
\usepackage{newfloat}
\usepackage[percent]{overpic}

\DeclareFloatingEnvironment[fileext=lop, listname={List of prompts}, 
                            name=Prompt, placement=h]{prompt}
\usepackage{listings}
\usepackage{xcolor}
\lstdefinelanguage{PolicyPrompt}{
  literate=
    {<think>}{{{\color{green!60!black}<think>}}}{7}
    {</think>}{{{\color{green!60!black}</think>}}}{8}
    {<search>}{{{\color{orange!80!black}<search>}}}{8}
    {</search>}{{{\color{orange!80!black}</search>}}}{9}
    {<information>}{{{\color{blue!80!black}<information>}}}{13}
    {</information>}{{{\color{blue!80!black}</information>}}}{14}
    {<answer>}{{{\color{red!70!black}<answer>}}}{8}
    {</answer>}{{{\color{red!70!black}</answer>}}}{9}
    {context\_block}{{{\color{red!90!black}context\_block}}}{13}
    {last\_user\_utterance}{{{\color{red!90!black}last\_user\_utterance}}}{19}
}

\usepackage{times}
\usepackage{latexsym}
\usepackage{amsmath}
\usepackage{amssymb}
\usepackage[T1]{fontenc}
\usepackage[utf8]{inputenc}
\usepackage{microtype}
\usepackage{inconsolata}
\usepackage{graphicx,xcolor}
\usepackage{booktabs}
\usepackage{pifont}

\usepackage[most]{tcolorbox}
\usepackage[table]{xcolor}
\usepackage{colortbl}

\newcommand{\header}[1]{\vspace{2mm}\noindent\textbf{#1}}

\acrodef{CS}{Conversational Search}
\acrodef{CSA}{Conversational Search Agent}
\acrodef{PTKB}{Personal Text Knowledge Base}
\acrodef{TREC}{TExt Retrieval Conference}
\acrodef{iKAT}{Interactive Knowledge Assistance Track}
\acrodef{CAsT}{Conversational Assistance Track}
\acrodef{NIST}{National Institute of Standards and Technology}
\acrodef{LLM}{Large Language Model}
\acrodef{LSR}{Learned Sparse Retrieval}

\acrodef{IR}{Information Retrieval}
\acrodef{NLP}{Natural Language Processing}
\acrodef{PEFT}{Parameter-Efficient Fine-Tuning}
\acrodef{ICL}{In-Context Learning}
\acrodef{LoRA}{Low-Rank Adaptation}

\acrodef{CQR}{Conversational Query Rewriting}

\acrodef{MSE}{Mean Square Error}

\newcolumntype{P}[1]{>{\centering\arraybackslash}p{#1}}

\usepackage{xcolor}
\usepackage{listings}
\lstdefinestyle{pythonstyle}{
    language=Python,
    basicstyle=\ttfamily\scriptsize,
    keywordstyle=\color{red!70!black}\bfseries,
    stringstyle=\color{green!50!black},
    commentstyle=\color{gray!70!black}\itshape,
    identifierstyle=\color{black},
    emph={f},
    emphstyle=\color{blue}\bfseries,
    numbers=none,
    breaklines=true,
    columns=fixed,
    basewidth=0.52em,
    keepspaces=true,
    showstringspaces=false,
    backgroundcolor=\color{gray!5},
    frame=single,
    framerule=0.3pt,
    rulecolor=\color{gray!20},
    framesep=2pt,
    xleftmargin=0pt,
    xrightmargin=0pt
}

\lstdefinestyle{javastyle}{
    language=Java,
    basicstyle=\ttfamily\scriptsize,
    keywordstyle=\color{red!70!black}\bfseries,
    stringstyle=\color{green!50!black},
    commentstyle=\color{gray!70!black}\itshape,
    identifierstyle=\color{black},
    emph={f},
    emphstyle=\color{blue}\bfseries,
    numbers=none,
    breaklines=true,
    columns=fullflexible,
    keepspaces=true,
    showstringspaces=false,
    backgroundcolor=\color{gray!5},
    frame=single,
    framerule=0.3pt,
    rulecolor=\color{gray!20},
    framesep=2pt,
    xleftmargin=0pt,
    xrightmargin=0pt
}

\lstdefinestyle{sqlstyle}{
    language=SQL,
    basicstyle=\ttfamily\scriptsize,
    keywordstyle=\color{red!70!black}\bfseries,
    stringstyle=\color{green!50!black},
    commentstyle=\color{gray!70!black}\itshape,
    identifierstyle=\color{black},
    emph={Student,Age,Group},
    emphstyle=\color{blue}\bfseries,
    numbers=none,
    breaklines=true,
    columns=fullflexible,
    keepspaces=true,
    showstringspaces=false,
    backgroundcolor=\color{gray!5},
    frame=single,
    framerule=0.3pt,
    rulecolor=\color{gray!20},
    framesep=2pt,
    xleftmargin=0pt,
    xrightmargin=0pt
}

\lstdefinestyle{instructionstyle}{
    language={},
    basicstyle=\ttfamily\scriptsize,
    numbers=none,
    breaklines=true,
    columns=fullflexible,
    keepspaces=true,
    showstringspaces=false,
    backgroundcolor=\color{gray!5},
    frame=single,
    framerule=0.25pt,
    rulecolor=\color{gray!20},
    framesep=2pt,
    xleftmargin=0pt,
    xrightmargin=0pt,
    aboveskip=2pt,
    belowskip=2pt,
    escapeinside={(*@}{@*)}
}

\title{On the Challenges and Opportunities of Learned \newline  Sparse Retrieval for Code}
\titlerunning{On the Challenges and Opportunities of Learned Sparse Retrieval for Code}

\correspondingauthor{s.c.lupart@uva.nl | $^{\dagger}${\bf Work done during internship at Naver Labs Europe}.}

\authors{Simon Lupart$^{\dagger}$ \authsep Maxime Louis \authsep Thibault Formal \authsep  Hervé Déjean \authsep Stéphane Clinchant}
\affiliations{NAVER LABS Europe}
\website{%
\includegraphics[height=1.2em]{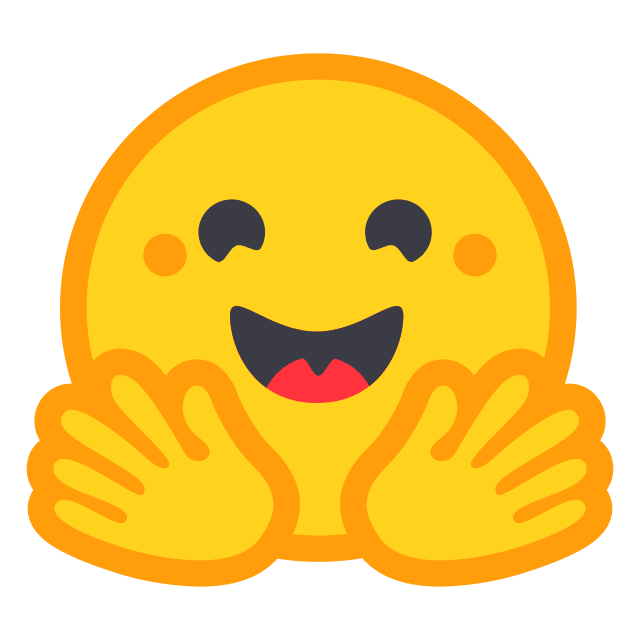}~\href{https://huggingface.co/naver/splade-code-06B}{naver/splade-code-0.6B}, 
\href{https://huggingface.co/naver/splade-code-8B}{naver/splade-code-8B}
}

\input{paper/0_abstract}

\begin{document}

\maketitle

\input{paper/1_intro}

\input{paper/2_related_work}

\input{paper/3_method}

\input{paper/4_experiments}

\input{paper/6_conclusion}

\clearpage
{
    \small
    \bibliographystyle{ieeenat_fullname}
    \bibliography{main}
}

\clearpage
\appendix
\input{supplementary/A_experiments}

\end{document}

%% file: paper/0_abstract.tex
\begin{abstract}
    Retrieval over large codebases is a key component of modern LLM-based software engineering systems. Existing approaches predominantly rely on dense embedding models, while learned sparse retrieval (LSR) remains largely unexplored for code. However, applying sparse retrieval to code is challenging due to subword fragmentation, semantic gaps between natural-language queries and code, diversity of programming languages and sub-tasks, and the length of code documents, which can harm sparsity and latency. We introduce SPLADE-Code, the first large-scale family of learned sparse retrieval models specialized for code retrieval (600M–8B parameters). Despite a lightweight one-stage training pipeline, SPLADE-Code achieves state-of-the-art performance among retrievers under 1B parameters (75.4 on MTEB Code) and competitive results at larger scales (79.0 with 8B). We show that learned expansion tokens are critical to bridge lexical and semantic matching, and provide a latency analysis showing that LSR enables sub-millisecond retrieval on a 1M-passage collection with little effectiveness loss.
\end{abstract}

%% file: paper/1_intro.tex
\section{Introduction}
\label{sec:introduction}

Large language models (LLMs) have recently achieved strong performance on a wide range of coding tasks~(\citeauthor{jiang2024survey,hui2024qwen2,khanfir2022codebert,li2023starcoder,olmo2025olmo}). Beyond standalone code generation, they are increasingly integrated into agentic frameworks~(\citeauthor{anthropic_claude_code_2026,mistral_vibe_2026}), where they interact with tools, repositories, and documentation to support software development workflows~(\citeauthor{santos2025decoding,yang2025code}). In these environments, retrieval plays a central role: models must explore large codebases, reuse existing components, and access external libraries or issue trackers such as GitHub~(\citeauthor{wang2025coderag}). Consequently, accurate and efficient code retrieval has become a fundamental building block of modern AI-driven software engineering systems. To the best of our knowledge, existing neural code retrieval approaches predominantly rely on \textit{dense} embedding models.

\begin{figure}[t!]
    \centering
    \begin{overpic}[width=\linewidth]{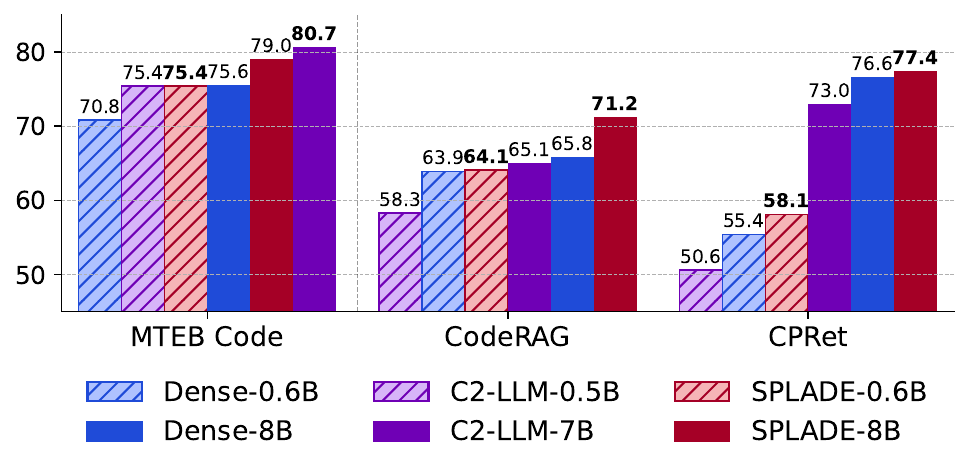}
        \put(12,48){\footnotesize In-domain}
        \put(54,48){\footnotesize Out-of-domain}
    \end{overpic}
    \label{fig:combined}
    \vspace{0.5em} 
    \input{figures/intro_lsr}
    \label{figure:intro}
\end{figure}

\header{Opportunities of LSR for code.} Learned sparse retrieval (LSR) models, which represent queries and documents as sparse high-dimensional bag-of-words vectors, may offer a compelling alternative for code retrieval. First, they often exhibit stronger generalization properties, being less sensitive to domain shifts and retrieval tasks~(\citeauthor{10.1145/3477495.3531857/splade++,lassance2024spladev3}).
Second, they usually offer low latency through the use of inverted index (\citeauthor{tonellotto2018efficient}) and their optimized implementations (\citeauthor{bruch2024efficient}). Finally, LSR models are also interpretable and thus easier to diagnose.

\header{Challenges of LSR.} However, building sparse retrieval models for code comes with its own challenges. \footnotetext{at the time of the submission, C2-LLM is the top 1 open source model on MTEB Code for 0.5B and 7B.}\textbf{(1)} First, code retrieval is heterogeneous: queries and documents may consist of natural language, code, or both, and code may come from different programming languages. As a result, the task requires strong text-to-code and code-to-code semantic alignment, akin to multilingual retrieval where semantics across languages has been challenging for LSR models~(\citeauthor{formal2026learning,nguyen2026milco}). 
\textbf{(2)} Second, most LSR models build sparse representations by projecting embeddings onto the fixed vocabulary of their backbone LLM~(\citeauthor{spladev1}), exposing them to fragmented sub-word splits that can affect their representations. These issues are particularly prominent on code (\citeauthor{li2025tokdrift}).
\textbf{(3)} Finally, LSR models are sensitive to the long input lengths. Since token activations accumulate across positions, longer inputs typically produce denser representations, increasing both index size and query latency. While this is manageable for short natural-language documents~(\citeauthor{10.1145/3477495.3531833/efficient}), code retrieval often involves long files or multi-function code snippets: even a classical BM25 incurs non-negligible latency (50 ms on a 1M collection in our setting). Without careful control of sparsity, LSR models may suffer from degraded efficiency at scale, making them impractical.

In this paper, we show how to tackle those challenges, achieving competitive results in effectiveness and efficiency for code retrieval. Our main contributions are as follows:
\begin{itemize}[leftmargin=10pt]
    \item We introduce SPLADE-Code, the first large-scale family of learned sparse retrieval models specialized for code, ranging from 600M to 8B parameters. SPLADE-code models can be trained with a lightweight one-stage training pipeline while still achieving state-of-the-art performance among retrievers under 1B parameters (75.4 on MTEB Code with 0.6B) and strong results at larger scales (79.0 with 8B). In line with the typical behavior of sparse retrieval models, we observe strong generalization to out-of-domain settings, as shown in Figure \ref{figure:intro} (top).

    \item We show that high performance critically relies on expansion tokens (i.e., tokens not present in the input), and that learned sparse representations combine strong exact lexical matching with robust semantic abstraction: example shown in Figure \ref{figure:intro} (bottom).

    \item We study the retrieval latency of LSR models for code, to evaluate their practicality in real-world settings. We achieve below 1 ms per query on the CodeSearchNet 1M passage collection using optimized LSR inverted-index retrieval, with only marginal ranking loss.
\end{itemize}

%% file: figures/intro_lsr.tex
\adjustbox{max width=\textwidth}{
\centering
\begin{minipage}{0.4\linewidth}
\footnotesize
\textbf{\textit{\underline{Example Code and LSR Representation:}}}
\vspace{0.4mm}
\begin{lstlisting}[style=pythonstyle]
def f(n):
    if n <= 1:
        return n
    return f(n-1) 
             + f(n-2)

x = int(input())
print(f(x))
\end{lstlisting}

\end{minipage}
\hfill
\begin{minipage}{0.48\linewidth}
\footnotesize
\centering
\vspace{6mm}

\begin{tabular}{l r}
\toprule
Token & Weight \\
\midrule
\cellcolor{red!9}\textit{Fibonacci}        & 2.45 \\
\cellcolor{red!9}\textit{fib}        & 2.22 \\
\cellcolor{red!9}\textit{recursive} & 2.12\\
\cellcolor{blue!9}input & 1.99 \\
\cellcolor{blue!9}f  & 1.90 \\
\cellcolor{red!9}\textit{Python} & 1.51 \\
\bottomrule
\end{tabular}
\end{minipage}
}
\caption{(Top) Performance in nDCG@10 of SPLADE-Code and dense models on code retrieval benchmark.\protect\footnotemark\ (Bottom) Example from CodeFeedback-ST.
Only a few vocabulary dimensions receive non-zero weights,
covering both \colorbox{red!9}{\textit{semantic}} terms (e.g., Fibonacci, Python, recursive)
and \colorbox{blue!9}{\textit{lexical}} tokens (e.g., input, f).}
\label{fig:fibo}

%% file: paper/2_related_work.tex
\section{Related Works}
\label{sec:related}

\header{Code retrieval tasks.} Code retrieval has recently attracted a lot of attention, leading to the development of several benchmarks and tasks~(\citeauthor{husain2019codesearchnet,suresh2024cornstack}). Applications cover Text-to-Code, Code-to-Text, Code-to-Code and Hybrid retrieval settings, including tasks such as bug fixing, code summarization, code completion, code translation, and programming contest problem solving, all framed as retrieval tasks~(\citeauthor{li-etal-2025-coir}). Current benchmarks also span an increasing number of programming languages, e.g. CPRet with more than 20 languages~(\citeauthor{deng2025cpret}), and domains, e.g. StackOverflow, GitHub Functions, Database, Code Instruction, Libraries~(\citeauthor{li-etal-2025-coir,wang2025coderag}). This requires retrieval models to be robust to domains and tasks, but also to have strong semantic capabilities.

\header{Code dense embedding.} A few recent works trained dense embedding models for code. One of the first, CodeXEmbed~(\citeauthor{liu2024codexembed}), proposed a multi-stage training LoRA using English text and code retrieval data, creating a specialized embedding model. CodeR-Pile~(\citeauthor{li2025towards}) proposed to synthesize a large-scale training dataset with a code-specialized LLM, adding more than 47 tasks and 20 programming languages, thus improving performance on benchmarks. More recently C2LLM~(\citeauthor{qin2025c2llm}) proposed a different pooling attention to reduce the information bottleneck in the EOS token. All these models have resource-intensive multi-stage training pipelines. 
Nevertheless, alternative representations remain largely unexplored for code retrieval; to the best of our knowledge, our work is the first to investigate learned sparse retrieval in this setting.

\header{Learned sparse retrieval} refers to a family of neural retrieval methods,
that learns to assign sparse vocabulary-based term weights to queries and documents, enabling semantic search with the efficiency and interpretability of inverted-index retrieval~(\citeauthor{spladev1,mallia2021learning,nguyen2023unified,SparseEmbed2023kong}). A seminal work has been the SPLADE models, showing excellent generalization properties~(\citeauthor{spladev1,10.1145/3477495.3531857/splade++,lassance2024spladev3,dejean_middletraining}).
SPLADE models, being based on BERT architectures have been extended with LLMs and decoder architectures~(\citeauthor{soares-etal-2023-nail,doshi2024mistralspladellmsbetterlearned, qiao2025leveragingdecoderarchitectureslearned, xu2025cspladelearnedsparseretrieval,10.1145/3726302.3730225,ma2025lightretrieverllmbasedhybridretrieval}), and also generalized to sparse auto-encoders~(\citeauthor{formal2026learning}). It has also been applied to cross-lingual and conversational search~(\citeauthor{nguyen2026milco,lupart25_disco}). Because code retrieval requires modeling both deep semantics and long contexts, it naturally aligns with SPLADE's ability to capture sparse rich semantic signals.

In addition, LSR enables the use of traditional inverted index structures to efficiently compute sparse dot-product scores~(\citeauthor{tonellotto2018efficient}). Similar to dense retrieval, additional approximate nearest neighbors methods have been developed~(\citeauthor{bruch2024efficient}), including posting list pruning~(\citeauthor{10.1145/3539618.3591941/pruning}) and sequential pruning~(\citeauthor{twostepsplade}) that could be studied for the long contexts of code retrieval tasks.

%% file: paper/3_method.tex
\section{Method}
\label{sec:method}

\subsection{Learned Sparse Retrieval}

Learned sparse retrieval (LSR) is a family of models that aim to create high-dimensional sparse representations, formally:
\begin{equation}
\mathbf{u}
=
\left(u_1,\ \dots,\ u_{N}\right)
\in \mathbb{R}^{N}, \quad N \in \mathbb{N^+}
\end{equation}
such that $\|\mathbf{u}\|_0 \ll {N}$. 
These sparse representations are used to compute similarity scores using a dot product $\mathbf{s}=\mathbf{u} \cdot \mathbf{v}$ between query and document representations.

Most high-performing LSR models are based on SPLADE~\cite{spladev1}. In SPLADE, sparse representations are obtained by projecting contextualized token representations onto the vocabulary $\mathcal V$ of the language model (LM), using its embedding matrix. Hence, $N=|\mathcal{V}|$: the LM vocabulary fixes the dimensionality. For an input sequence of length $n$, sparse vectors are also aggregated across the sequence length with a max pooling:
\begin{equation}
\mathbf{u}_t
=
\max_{i \in \{1,\ldots,n\}}
\log(1+\mathrm{ReLU}(\mathbf{z}_{i,t}))
\end{equation}
where $\mathbf{z}_{i,t}$ denotes the logit associated with token $t$ at position $i$, after the projection to the vocabulary space. 
This max pooling operation selects, for each vocabulary term, the most salient activation across the sequence, yielding a sparse bag-of-words–like representation. The $\log(1 + \mathrm{ReLU}(\cdot))$ activation ensures non-negative features and induces sparsity.

\subsection{SPLADE for Code Retrieval}

\begin{figure}[t!]
    \centering
    \includegraphics[width=\linewidth]{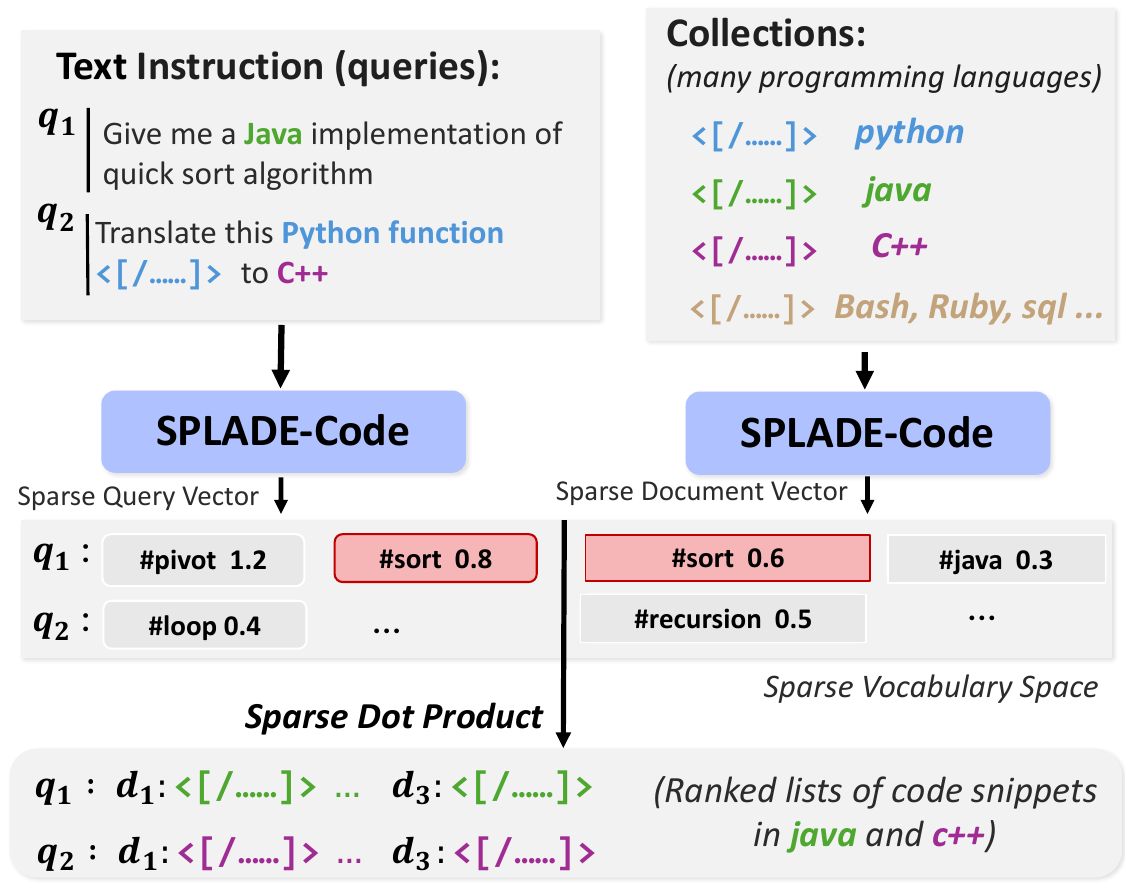}
    \caption{SPLADE-Code is a bi-encoder model for retrieval across programming languages (e.g. Text2Code, Code2Code). When queries and documents are in different programming languages, SPLADE-Code learn some shared dimensions in the representation space.}
    \label{fig:densevslsr}
\end{figure}

\header{SPLADE-Code.}
As illustrated in Figure~\ref{fig:densevslsr}, SPLADE-Code is an LSR model for code retrieval. Unlike standard text retrieval, code retrieval requires the model to understand both textual query descriptions and structured source code. The model must thus capture semantic intent (e.g., sorting, parsing, recursion, database queries) while remaining robust to syntactic variation and programming language differences.

\header{Training objective for code.}
To train the models for the code retrieval task, we use a Kullback-Leibler divergence (KLD) distillation objective~(\citeauthor{Kullback1951OnIA,lin2020distilling}). This objective is widely adopted for LSR and dense text retrieval, to distil relevance scores from cross-encoder teacher models to retrieval models~(\citeauthor{hofstatter2020improving,10.1145/3477495.3531857/splade++}).
Formally, let $q$ be a query and $\{d_1,\ldots,d_{K+1}\}$ a candidate set of one positive document and $K$ mined negatives in any programming language. The KLD loss is computed between the distributions of similarity scores obtained from the teacher and the student models, on the same set of $(q,d)$ pairs:
\begin{equation}
\mathcal{L}_{\mathbf{KLD}} = \mathcal{D}_{\mathbf{KL}}(\mathbf{S}_{T} \,||\, \mathbf{S}_{S})
\label{eq:kld}
\end{equation}
where $\mathbf{S}_{S}$ and $\mathbf{S}_{T}$ are the softmax of scores distribution of student and teacher models. This objective encourages the student model to match the teacher's relative ranking over the candidate documents, rather than relying solely on hard relevance labels. Because queries and documents may appear in different programming languages or as textual descriptions, the objective encourages the model to project all inputs into shared sparse dimensions.

In addition, following SPLADE, and to encourage sparse activations, we also apply a FLOPs penalty within each batch. This regularizer penalizes terms that are frequently activated across the batch, which empirically leads to sparser posting lists and lower retrieval cost~(\citeauthor{paria2020minimizing}).

%% file: paper/4_experiments.tex
\subsection{Model Configuration}

\header{Hyper-parameter details.} We train our model on CoIR (see Table~\ref{stat_mini}) with LoRA~(\citeauthor{hu2022lora}), a batch size of 256 with 7 negatives per query, and a max length of 512. We use a temperature of 300 in the KLD loss, with a learning rate of 1e-4.
The negatives were mined from retrieval made with Qwen3-Embedding-0.6B, and scored with  Qwen3-Reranker-4B to be then distilled~(\citeauthor{zhang2025qwen3}). We select the negative between ranks 50 and 100. More details are provided in Appendix~\ref{sec:appendix:hyperparam}.

\header{Bi-directional attention.} Since standard decoder-only models use causal (uni-directional) self-attention, directly applying them restricts each token's representation to left context only, which is suboptimal for lexical matching~(\citeauthor{xu-etal-2025-csplade}).
To address this, we enable bidirectional self-attention and pre-train them on MS MARCO passages~(\citeauthor{msmarco}) with Masked Next Token Prediction, following prior work on LSR~(\citeauthor{qiao2025causalsplade,hansi2025causalsplade}).

\header{Model merging} is done with weighted spherical merging~(\citeauthor{yang2024modelmerging}) from three checkpoints: (1) the base model after the first epoch, (2) one model after a second epoch, (3) one model trained with a max context length of 1024 for one epoch. Small models (0.6B) also contain a fourth checkpoint trained with full finetuning (no LoRA).

\section{Experiments}
\label{sec:exps}

\subsection{Benchmarks and Evaluation}
\label{sec:expset}

We evaluate our models on various code retrieval benchmarks, containing different programming languages, tasks and characteristics. For \textit{in-domain} evaluation, we rely on:
\begin{itemize}[leftmargin=10pt]
    \item CoIR~(\citeauthor{li-etal-2025-coir}): the biggest code retrieval benchmark, containing ten datasets in 14 code languages. It features 4 main tasks, Code2Code, Text2Code, Code2Text and Hybrids with train/test splits. All merged training datasets contain 2.2M training samples.
    \item MTEB-code~(\citeauthor{muennighoff-etal-2023-mteb}): built upon CoIR, with two new datasets: CodeEditSearch Retrieval and new variation of CodeSearchNet. We consider it in-domain since it is mainly composed of CoIR datasets.
\end{itemize}
\noindent
In addition, we use two additional benchmarks for \textit{out-of-domain} evaluation, namely:
\begin{itemize}[leftmargin=10pt]
    \item CodeRAG~(\citeauthor{wang2025coderag}): a retrieval-augmented code generation benchmark, containing retrieval annotation on 8 coding tasks: basic programming, open-domain, and repository-level problems.
    \item CPRet~(\citeauthor{deng2025cpret}): a benchmark containing 4 retrieval datasets focusing on problem-centric competitive programming with more than 20 programming languages.
\end{itemize}
\noindent
We report the main retrieval metrics, using nDCG@10~(\citeauthor{ndcg}) following common practice on the used benchmarks. This evaluates the top ranking of each retrieved ranked list using the relevance annotation from evaluated datasets.
For the LSR retrieval we used the \textit{seismic} library, which contains an efficient implementation of inverted index~(\citeauthor{bruch2024efficient}).

\input{tables/dataset_short}

\header{Baselines.}
We compare SPLADE-Code against dense retrievers trained with the same pipeline on Qwen3-0.6B, 1.7B, and 8B, as well as top models from the MTEB Code leaderboard. These include C2-LLM~(\citeauthor{qin2025c2llm}), CodeXEmbed~(\citeauthor{liu2024codexembed}) and CodeR, aka bge-code-v1~(\citeauthor{li2025towards}), trained on CoIR and synthetic Qwen2.5Coder-32B data.
We further compare SPLADE-Code to other LSR variants, including SPLADE-lexical, which uses token weighting without expansion, BM25, alternative LLM backbones (QwenCoder, Llama), and SPLARE, which learns sparse features via sparse autoencoders.

\subsection{Main Results}

\input{tables/codellm_dense}

\input{tables/mteb_code_main_us}

\input{tables/code_rag_main}

\header{LSR and dense in controlled experiments.}
Table~\ref{tab:main} presents the performance of LSR and dense models on four code retrieval benchmarks under controlled experimental settings. All models are trained on the CoIR training split using comparable hyperparameters and similar backbone architectures to ensure a fair comparison (cf Appendix \ref{sec:appendix:hyperparam}). The evaluation spans 22 datasets and covers more than 20 programming languages across a variety of subtasks.

Dense models achieve high performance overall, with nDCG@10 scores typically in the 0.70-0.80 range, with consistent improvements as the backbone model size increases, providing strong baselines. Yet, our results show that SPLADE-Code consistently matches or exceeds dense baselines across the four benchmarks. Compared to dense, SPLADE-Code-8B outperforms them in both in-domain and out-of-domain evaluation. These improvements also remain stable across different model scales.

In contrast, purely lexical approaches such as SPLADE-lexical perform poorly. These models lack the semantic expansion required to capture the meaning of code-related queries and documents. Similarly, BM25 struggles in this setting for the same reason. This behavior is expected, as code retrieval tasks are highly semantic, particularly for Text-to-Code and Code-to-Code retrieval across different programming languages.

\header{In-domain performance comparison with top models.}
Table~\ref{tab:mteb-code} compares SPLADE-Code with state-of-the-art dense retrievers. In this setting, we apply checkpoint merging to improve the LSR models, which yields gains of approximately 1-2 nDCG@10 points compared to the original SPLADE-Code models reported in Table~\ref{tab:main}. We also focus on the comparison with C2-LLM, which represents the closest experimental setting to ours (similar sizes, and both trained on CoIR using checkpoint merging), in contrast to CodeR-all, which was trained on 5.2M training samples (vs. 2.2M for all other models).

For smaller models, SPLADE-Code-0.6B shows state-of-the-art performance among systems below one billion parameters. In particular, it achieves comparable results to C2-LLM-0.5B, and ranks first on CoIR within this model size category.
Now at larger scales, SPLADE-Code-8B achieves the best performance on three datasets (CodeSearchNet, CodeTrans-Contest, and Stack- Overflow-QA) among the systems considered, suggesting that certain code retrieval tasks benefit from LSR over dense retrieval. Compared with C2-LLM-7B and CodeR-all, SPLADE-Code-8B obtains particularly lower performance on CosQA and CodeTrans-DL. This difference may be partially explained by dataset characteristics, as CosQA has been shown to exhibit strong annotation biases~(\citeauthor{cosqa++}). Dense models may more easily adapt to such biases, whereas LSR models rely on vocabulary-based sparse representations, which may make fitting these dataset-specific patterns more challenging.

\header{Generalization to out-of-domain benchmarks.}
Table~\ref{tab:coderag_cpret_merged} reports a detailed comparison with top models on the out-of-domain CodeRAG Bench and CPRet benchmarks. In this setting, SPLADE-Code achieves higher performance than most models. In particular, it surpasses C2-LLM by 5.8 and 6.1 nDCG@10 points on CodeRAG Bench for both the 0.6B and 8B configurations, and by 7.5 and 4.4 points on CPRet, respectively.
Datasets such as DS-1000, ODEX, T2C, and C2C illustrate the limitations of lexical-only retrieval. On these datasets, BM25 achieves nDCG@10 values typically below 10, where in contrast, SPLADE-Code has +30 nDCG points gains. This behavior suggests that LSR reduces lexical mismatch and captures semantic meaning between queries and code.

More broadly, the strong out-of-domain performance aligns with prior findings on SPLADE models, which have shown robust generalization across domains~(\citeauthor{lassance2024spladev3}). Because sparse representations are grounded in the backbone vocabulary, they can remain effective even when queries and documents differ from the training distribution, as the model can still rely on lexical and shared semantic terms within the vocabulary space.
Finally, SPLADE-Code remains competitive with CodeR-all, which was trained on a much larger dataset. Since some evaluation datasets overlap with CodeR-all training data, its evaluation is not strictly out-of-domain. Despite this advantage, SPLADE-Code achieves comparable performance with a substantially lighter training setup.

\subsection{Analysis}

\begin{figure}[t]
    \centering
    \includegraphics[width=\linewidth]{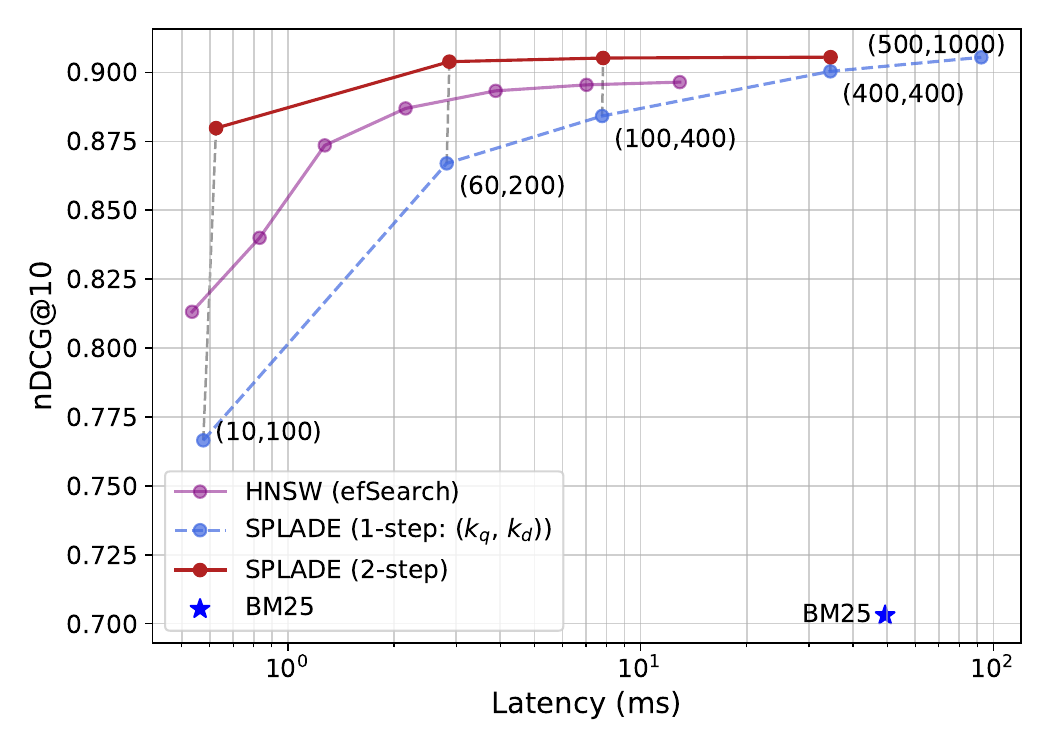}
    \caption{Effectiveness--efficiency trade-off on the CodeSearchNet 1M passage collection. The figure reports retrieval latency and effectiveness of SPLADE-Code-8B compared to the Dense-8B model. $(k_q, k_d)$ indicates the pruning used on queries and documents for LSR.}
    \label{fig:latency}
\end{figure}

\header{Effectiveness--Efficiency trade-off.}
Figure~\ref{fig:latency} shows the effectiveness–efficiency trade-off of SPLADE-Code, reporting retrieval latency in milliseconds per query against retrieval effectiveness (average latency on the 53k test queries). 
Overall, SPLADE-Code achieves efficient retrieval through pruning of the sparse representations. We evaluate a two-step SPLADE retrieval strategy~(\citeauthor{twostepsplade}), where an initial stage applies aggressive pruning on query and document representations, e.g. $(10, 100)$, to quickly retrieve a candidate set, followed by a second stage that refines the scores with a less aggressive pruning configuration, in practice $(500, 1000)$. Because both stages rely on the same sparse representations, this approach preserves most of the retrieval effectiveness while substantially reducing latency.
In contrast, BM25 exhibits high latency (49\,ms per query\footnote{using pyserini~(\citeauthor{Lin_etal_SIGIR2021_Pyserini}) BM25 inverted index.}), due to the long average lengths of queries and documents in CodeSearchNet (594 and 156 average words respectively), which results in a large number of postings to process in the inverted index.
For comparison, dense retrieval uses the HNSW algorithm~(\citeauthor{hnsw2020malkov}) implemented in FAISS~(\citeauthor{douze2025faiss}), and all experiments are conducted on the same single AMD EPYC 7313 processor, while LSR uses the \textit{Seismic} inverted index library.

\input{tables/examples}

\header{Effect of the projection space.}
Table~\ref{tab:backbone} compares LSR models built on different backbone LLMs, which define the projection space. Overall, performance is relatively stable across 7-8B backbones. In particular, generalist models such as Qwen3-8B, Llama3-8B, and code-specialized backbones like Qwen2.5Coder-7B yield nearly identical results across benchmarks. Comparing with smaller models, on Qwen2.5Coder-0.5B and Qwen2.5Coder-1.5B, results are also similar to those for models trained on Qwen3-0.6B and Qwen3-1.7B. SPLADE-Code seems to be robust to the projection vocabulary.
Finally, replacing vocabulary-based projection with learned sparse latent features, as in SPLARE (\citeauthor{formal2026learning}), produces comparable results. This may be because the matching space remains largely grounded in English natural language, which acts as a shared pivot between queries and documents. As a result, differences in the projection vocabulary appear to have limited impact on retrieval effectiveness.

\header{Training pipeline variants.}
Table~\ref{tab:lsr_ablation_code} compares several training variants. Both the base setup and the instruction-tuned variant achieve similar performance, indicating that SPLADE-Code works well with or without instructions in this setting. 

In contrast, intermediate English fine-tuning leads to lower effectiveness on both CoIR and MTEB Code. This differs from dense retrieval, where an intermediate text retrieval stage is often beneficial. A plausible explanation is that LSR already retains lexical matching capabilities through its sparse representations, reducing the need for an English intermediate stage before training on code.

\header{Interpretability of the LSR representations.}
A key advantage of SPLADE-style LSR is interpretability: the representations are highly sparse, with only a few hundred active dimensions in a vocabulary-sized space, and each dimension can be directly mapped to a token.
Figure~\ref{fig:examples} illustrates this behavior with three examples.

\input{tables/backbone}

\input{tables/ablations}

In Example~1 (Python image loading), the model expands the representation with semantically related terms such as \texttt{pillow} and \texttt{PIL}, as well as exception types (e.g., \texttt{IOError}, \texttt{FileNotFoundError}). These activations align with common failure modes and help retrieve passages describing similar issues. 
In Example~2 (SQL query), the model downweights generic syntactic tokens (e.g., \texttt{SELECT}, \texttt{FROM}, \texttt{WHERE}) and instead emphasizes content-bearing terms such as \texttt{oldest} from \texttt{MAX(Age)}. This suggests that the representation prioritizes semantics over syntax. 
In Example~3 (Java implementation of quicksort), the input code does not explicitly mention ``quicksort'', yet the model activates algorithmic and implementation-level concepts such as \texttt{quick}, \texttt{pivot}, \texttt{sort}, and \texttt{java}, along with related terms like \texttt{swap}, \texttt{inplace}, and \texttt{array}. Such activations shows how SPLADE-Code uses mostly English as matching language between textual instructions and code from different programming languages.

More broadly, the activated terms remain semantically coherent, and we do not observe obvious concept drift toward noisy or non-semantic tokens~(\citeauthor{nguyen2024multimodal}). Among the top-25 activated terms, roughly 35\% come from the input itself and 65\% are expansion terms, highlighting the central role of expansion in the learned sparse representation. For clarity in the examples provided, we also omitted a few lexical tokens with different capitalization or subword tokens from the top weights.

%% file: tables/dataset_short.tex
\begin{table}[t]
\centering
\small
\adjustbox{max width=0.49\textwidth}{
\begin{tabular}{l c}
\toprule
{Benchmark} & {Summary} \\
\midrule

\textbf{CoIR (Code IR)} & 10 datasets, 14 programming langs \\
 (\citeauthor{li-etal-2025-coir}) & \textbf{Train: 561-905K, Total: 2.2M} \\
& Test queries: 180--53K \\
\colorbox{gray!10}{(in-domain)} & Collection: 816--1M passages\\

\midrule

\textbf{MTEB-Code} & 12 datasets: 10 CoIR + 2 new \\
 (\citeauthor{muennighoff-etal-2023-mteb}) & 14 programming langs \\
 \textit{(build upon CoIR)}& Test queries: 180--53K \\
\colorbox{gray!10}{(in-domain)}  & Collection: 816--1M passages \\

\midrule

\textbf{CodeRAG-Bench}   & 6 datasets, 1 lang (Python) \\
 (\citeauthor{wang2025coderag})& Test queries: 164--1K \\
\colorbox{gray!10}{(out-of-domain)} & Collection: 237--40.9K passages \\

\midrule

\textbf{CPRet} & 4 datasets, 20 programming langs \\
 (\citeauthor{deng2025cpret})& Test queries: 168--10K \\
\colorbox{gray!10}{(out-of-domain)} & Collection: 10K--41.6K passages \\

\bottomrule
\end{tabular}
}

\caption{Benchmark overview. Sizes are reported as min--max numbers of samples per dataset (See Appendix \ref{sec:appendix:stat} for statistics and programming langs details).}
\label{stat_mini}
\end{table}

%% file: tables/codellm_dense.tex
\begin{table}[t]
\centering
\adjustbox{max width=0.49\textwidth}{
\begin{tabular}{l|cccc}
\toprule
\multirow{2}{*}{Model}  &\multirow{2}{*}{\textbf{CoIR}} & \textbf{MTEB} & \textbf{Code} & \multirow{2}{*}{\textbf{CPRet}}  \\
     &    &   \textbf{Code} &   \textbf{RAG}  &      \\
\midrule
\multicolumn{3}{l}{\textit{\textbf{Dense Retrieval}}} & & \\
Dense-0.6B & 71.4 & 70.8 & 63.9 & 55.4 \\
Dense-1.7B  & 73.4 & 73.4 & \underline{67.3} & 65.8 \\
Dense-8B  & \underline{76.0} & \underline{75.6} & 65.8 & \underline{76.6} \\
\midrule
\multicolumn{3}{l}{\textit{\textbf{Lexical Retrieval (no expansion)}}} & & \\ 
BM25  & 45.8 & 47.9 & 57.7 & 18.6 \\
SPLADE-lex.-8B & 46.2 & 45.8 & 56.7 & 29.4 \\
\midrule
\multicolumn{3}{l}{\textit{\textbf{Learned Sparse Retrieval}}} & & \\
SPLADE-Code-0.6B  & 72.6  & 73.5 & 63.7 & 56.4 \\
/vs Dense 
& {\color{green!60!black}(\textbf{+1.2})}
& {\color{green!60!black}(\textbf{+2.7})}
& {\color{red!70!black}({-0.2})}
& {\color{green!60!black}(\textbf{+1.0})} \\
SPLADE-Code-1.7B & 74.3 & 75.1 & 64.9 & 62.5 \\
/vs Dense  
& {\color{green!60!black}(\textbf{+0.9})}
& {\color{green!60!black}(\textbf{+1.7})}
& {\color{red!70!black}({-2.4})}
& {\color{red!70!black}({-3.3})} \\
SPLADE-Code-8B & \textbf{76.7} & \textbf{77.8} & { \textbf{68.5}} & \textbf{76.8} \\
/vs Dense & {\color{green!60!black}(\textbf{+0.7})}
& {\color{green!60!black}(\textbf{+2.2})}
& {\color{green!60!black}(\textbf{+2.7})} 
& {\color{green!60!black}(\textbf{+0.2})} \\
\bottomrule
\end{tabular}
}
\caption{Controlled experiments comparing learned sparse retrieval, dense models, and lexical baselines using nDCG@10. All models are trained on the same data without checkpoint merging to ensure a fair comparison.}
\label{tab:main}
\end{table}

%% file: tables/mteb_code_main_us.tex
\begin{table*}[t!]
\centering
\adjustbox{max width=\textwidth}{
\begin{tabular}{lcccccccccccc|cc}
\toprule
\multicolumn{1}{c}{\multirow{2}{*}{Model}} &
\multicolumn{1}{c}{\multirow{2}{*}{Apps}} &
\multicolumn{1}{c}{\multirow{2}{*}{CosQA}} &
\multicolumn{1}{c}{\multirow{2}{*}{T2SQL}} &
\multicolumn{1}{c}{\multirow{2}{*}{CSN$^\dagger$}} &
\multicolumn{1}{c}{\multirow{1}{*}{CSN}} &
\multicolumn{1}{c}{\multirow{1}{*}{CSN}} &
\multicolumn{1}{c}{\multirow{1}{*}{Edit}} &
\multicolumn{2}{c}{\multirow{1}{*}{CodeTrans}} &
\multicolumn{1}{c}{\multirow{1}{*}{SO}} &
\multicolumn{2}{c}{Feedback} &
\multicolumn{1}{c}{\multirow{1}{*}{\textbf{Avg}}} & 
\multicolumn{1}{c}{\multirow{1}{*}{\textbf{Avg}}} \\
&
& & & & -COIR & -CCR &
Ret.$^\dagger$ &
-CT &
-DL &
QA &
-ST &
-MT &
\textbf{COIR} &
\textbf{MTEB}
\\
\midrule
BM25 & 4.7 & 18.7 & 24.9 & 63.5 & 70.3 & 59.3 & 53.3  & 47.7 & {34.4} & 70.2 & 68.1 & 59.1 & \cellcolor{gray!15}45.8 & \cellcolor{gray!15}47.9 \\
\midrule
\multicolumn{3}{l}{\textit{\textbf{SotA Dense Retrieval}}} & & & & & && & & & & & \\
Jina-v2-code & 16.3 & 41.0 & 44.2 & -- & 84.0 & 82.7 & -- & 86.6 & 30.5 & 89.4 & 69.0 & 52.1 & \cellcolor{gray!15}59.6 & \cellcolor{gray!15}-- \\
Dense-0.6B & 51.8 & 33.3 & 71.1 & 77.5 & 85.8 & 92.0 & 58.7 & 86.2 & {32.6} & 91.5 & 80.2 & 89.4 & \cellcolor{gray!15}71.4 & \cellcolor{gray!15}70.8 \\
Dense-8B & 82.1 & 32.2 & {72.2} & 78.9	 & 88.9 & {95.2} & {68.6} & {93.1} & 28.2 & 95.8 &  {80.6} & {91.6} & \cellcolor{gray!15}{76.0} & \cellcolor{gray!15}{75.6}  \\
CodeR-all-1.5B & \textbf{98.1} & \underline{46.7} & 64.4 & 90.5 & 89.5  & \textbf{98.3} & 70.8 & 94.4 & \underline{46.1} & 95.4 & \underline{90.6} & \textbf{94.4} & \cellcolor{gray!15}\textbf{81.8} & \cellcolor{gray!15}\textbf{81.1} \\
CodeR-coir-1.5B & 81.5 & \textbf{46.6} & 64.2 & -- & 89.1  & {97.8} & -- & 94.4 & \textbf{46.3} & 95.1 & {89.6} & {93.2} & \cellcolor{gray!15}\underline{79.8} & \cellcolor{gray!15}-- \\
CodeXEmb-2B & 76.9 & 40.5 & \underline{78.4} & --& 87.9  & 97.7 & -- & 90.3 & 38.6 & {94.5} & 86.4 & 65.5 & \cellcolor{gray!15}75.7 & \cellcolor{gray!15}-- \\
CodeXEmb-7B & 85.4 & {42.5} & \textbf{78.9} & --& \underline{89.7}  &\underline{ 98.0} & -- & \underline{94.5} & {40.5} & \underline{96.3} & 87.5 & 68.8 & \cellcolor{gray!15}78.2 &\cellcolor{gray!15} -- \\
C2-LLM-0.5B & 61.0 & 38.3 & 74.1 & 89.2 & 86.7 & 96.3 & 71.4 & 84.3 & 34.0 & 89.4 & 88.6 & 92.3 & \cellcolor{gray!15}72.6 & \cellcolor{gray!15}75.4 \\
C2-LLM-7B & \underline{86.7} & 39.8 & 75.8 & 91.1 & \textbf{89.8 }& 97.9 & \textbf{81.5} & 92.5 & 34.1 & 94.9 & \textbf{90.7} & \underline{94.3} & \cellcolor{gray!15}{79.7} & \cellcolor{gray!15}\underline{80.7} \\
\midrule
\multicolumn{5}{l}{\textit{\textbf{Learned Sparse Retrieval}}}  & & & && & & & & & \\
SPLADE-Code-0.6B & 66.5 & {35.4} & 74.3 & 90.5 & 86.2 & 91.4 & 68.4 & 90.9 & 31.4 & 93.3 & 84.7 & 92.0 & \cellcolor{gray!15}74.6 & \cellcolor{gray!15}75.4 \\
/vs C2-LLM-0.5B
& {\color{green!60!black}(\textbf{+5.5})}
& {\color{red!70!black}({-2.9})}
& {\color{green!60!black}(\textbf{+0.2})}
& {\color{green!60!black}(\textbf{+1.3})}
& {\color{red!70!black}({-0.5})}
& {\color{red!70!black}({-4.9})}
& {\color{red!70!black}({-3.0})}
& {\color{green!60!black}(\textbf{+6.6})}
& {\color{red!70!black}({-2.6})}
& {\color{green!60!black}(\textbf{+3.9})}
& {\color{red!70!black}({-3.9})}
& {\color{red!70!black}({-0.3})}
& \cellcolor{gray!0}{\color{green!60!black}(\textbf{+2.0})}
& \cellcolor{gray!0}{\color{green!60!black}(\textbf{+0.0})} \\
SPLADE-Code-8B & \underline{86.7} & 32.5 & {75.7} & \textbf{92.1} &  88.9 & 94.6 & \underline{77.1} & \textbf{94.6} & 28.1 &  \textbf{96.5} & {87.6} & {93.9} & \cellcolor{gray!15}77.9 & \cellcolor{gray!15}{79.0} \\
/vs C2-LLM-7B
& {\color{green!60!black}(\textbf{+0.0})}
& {\color{red!70!black}({-7.3})}
& {\color{red!70!black}({-0.1})}
& {\color{green!60!black}(\textbf{+1.0})}
& {\color{red!70!black}({-0.9})}
& {\color{red!70!black}({-3.3})}
& {\color{red!70!black}({-4.4})}
& {\color{green!60!black}(\textbf{+2.1})}
& {\color{red!70!black}({-6.0})}
& {\color{green!60!black}(\textbf{+1.6})}
& {\color{red!70!black}({-3.1})}
& {\color{red!70!black}({-0.4})}
& \cellcolor{gray!0}{\color{red!70!black}({-1.8})}
& \cellcolor{gray!0}{\color{red!70!black}({-1.7})} \\

\bottomrule
\end{tabular}
}
\caption{MTEB-Code retrieval results (nDCG@10). All results are reported with pruning $(500,1000)$. Datasets included in MTEB-Code but not in CoIR are marked with $^\dagger$.}
\label{tab:mteb-code}
\vspace{0.5em}
\end{table*}

%% file: tables/code_rag_main.tex
\begin{table*}[t]
\centering
\adjustbox{max width=\textwidth}{
\begin{tabular}{lccccccc|ccccc}
\toprule

\multirow{2}{*}{Model} 
& \multicolumn{7}{c}{\textbf{CodeRAG-Bench}} 
& \multicolumn{5}{c}{\textbf{CPRet}} \\

\cmidrule(r){2-8} \cmidrule(l){9-13}

& Human & \multirow{2}{*}{MBPP} & \multirow{2}{*}{DS-1K} & \multirow{2}{*}{ODEX} 
& Repo & SWE & \multirow{2}{*}{\textbf{Avg}}
& \multirow{2}{*}{T2C} & \multirow{2}{*}{C2C} 
& \multirow{2}{*}{P2Du} & \multirow{2}{*}{S2Fu} 
& \multirow{2}{*}{\textbf{Avg}} \\

& Eval &  &  &  & Eval & Ben-L &  &  &  &  &  &  \\

\midrule
BM25 
& 100.0 & 98.6 & 5.2 & 6.7 & 93.2 & 43.0 & \cellcolor{gray!15}57.7
& 0.9 & 7.4 & 12.4 & 53.8 & \cellcolor{gray!15}18.6 \\

\midrule
\multicolumn{7}{l}{\textit{\textbf{SotA Dense Retrieval}}} & \cellcolor{gray!0} & & & & & \cellcolor{gray!0}\\

Dense-0.6B 
& 100.0 & 99.1 & 29.0 & 16.9 & 92.1 & 46.4 & \cellcolor{gray!15}63.9
& 32.3 & 43.7 & 54.2 & 91.4 & \cellcolor{gray!15}55.4 \\

Dense-8B 
& 100.0 & 98.9 & 30.3 & 26.4 & 89.0 & 50.4 & \cellcolor{gray!15} 65.8
& 64.3 & 74.2 & \textbf{72.5} & \underline{95.3} & \cellcolor{gray!15}76.6 \\

CodeR-all-1.5B
& 100.0 & \underline{99.2} & \underline{40.8} & 36.1 & 93.1 & \textbf{67.4} & \cellcolor{gray!15}\textbf{72.8}
& \textbf{75.1} & \textbf{85.4} & 58.4 & 94.7 & \cellcolor{gray!15}\textbf{78.3} \\

CodeR-coir-1.5B
& 100.0 & 98.9 & 37.2 & 32.5 & 92.2 & \underline{64.6} & \cellcolor{gray!15}70.9
& -- & -- & -- & -- & \cellcolor{gray!15}-- \\

CodeXEmbed-2B
& 100.0 & 97.4 & 25.4 & 23.9 & 88.7 & 52.4 & \cellcolor{gray!15}64.6
& 39.6 & 68.0 & 45.3 & 86.4 & \cellcolor{gray!15}59.8 \\

SFR-mistral-7B
& 100.0 & 99.0 & 19.3 & \underline{37.1} & 83.8 & 62.7 & \cellcolor{gray!15}67.0
& 22.2 & 50.9 & 31.9 & 69.4 & \cellcolor{gray!15}43.6 \\

C2-LLM-0.5B
& 99.8 & 98.6 & 18.1 & 21.0 & 67.6 & 44.6 & \cellcolor{gray!15}58.3
& 38.1 & 50.6 & 31.3 & 82.4 & \cellcolor{gray!15}50.6 \\

C2-LLM-7B
& 100.0 & 99.1 & 34.7 & 31.2 & 72.3 & 53.2 & \cellcolor{gray!15}65.1
& 67.8 & 75.6 & 54.6 & 94.0 & \cellcolor{gray!15}73.0 \\

\midrule
\multicolumn{7}{l}{\textit{\textbf{Learned Sparse Retrieval}}} & \cellcolor{gray!0} & & & & & \cellcolor{gray!0}\\

SPLADE-Code-0.6B
& 100.0 & \underline{99.2} & 21.1 & 29.1 & \textbf{95.8} & 39.5 & \cellcolor{gray!15}64.1
& 42.4 & 47.6 & 49.7 & 92.6 & \cellcolor{gray!15}58.1 \\
/vs C2-LLM-0.5B
& {\color{green!60!black}(\textbf{+0.2})}
& {\color{green!60!black}(\textbf{+0.6})}
& {\color{green!60!black}(\textbf{+3.0})}
& {\color{green!60!black}(\textbf{+8.1})}
& {\color{green!60!black}(\textbf{+28.2})}
& {\color{red!70!black}({-5.1})}
& \cellcolor{gray!0}{\color{green!60!black}(\textbf{+5.8})}
& {\color{green!60!black}(\textbf{+4.3})}
& {\color{red!70!black}({-3.0})}
& {\color{green!60!black}(\textbf{+18.4})}
& {\color{green!60!black}(\textbf{+10.2})}
& \cellcolor{gray!0}{\color{green!60!black}(\textbf{+7.5})} \\

SPLADE-Code-8B
& 100.0 & \textbf{99.4} & \textbf{42.0} & \textbf{38.8} & \underline{94.2} & 52.6 & \cellcolor{gray!15}\underline{71.2}
& \underline{70.9} & \underline{76.4} & \underline{66.3} & \textbf{96.1} & \cellcolor{gray!15}\underline{77.4} \\
/vs C2-LLM-7B
& {\color{green!60!black}(\textbf{+0.0})}
& {\color{green!60!black}(\textbf{+0.3})}
& {\color{green!60!black}(\textbf{+7.3})}
& {\color{green!60!black}(\textbf{+7.6})}
& {\color{green!60!black}(\textbf{+21.9})}
& {\color{red!70!black}({-0.6})}
& \cellcolor{gray!0}{\color{green!60!black}(\textbf{+6.1})}
& {\color{green!60!black}(\textbf{+3.1})}
& {\color{green!60!black}(\textbf{+0.8})}
& {\color{green!60!black}(\textbf{+11.7})}
& {\color{green!60!black}(\textbf{+2.1})}
& \cellcolor{gray!0}{\color{green!60!black}(\textbf{+4.4})} \\
\bottomrule
\end{tabular}
}
\caption{Out-of-domain retrieval performance on CodeRAG and CPRet (nDCG@10). All results are reported with pruning $(1000,1000)$.}
\label{tab:coderag_cpret_merged}
\end{table*}

%% file: tables/examples.tex
\begin{figure*}[t]
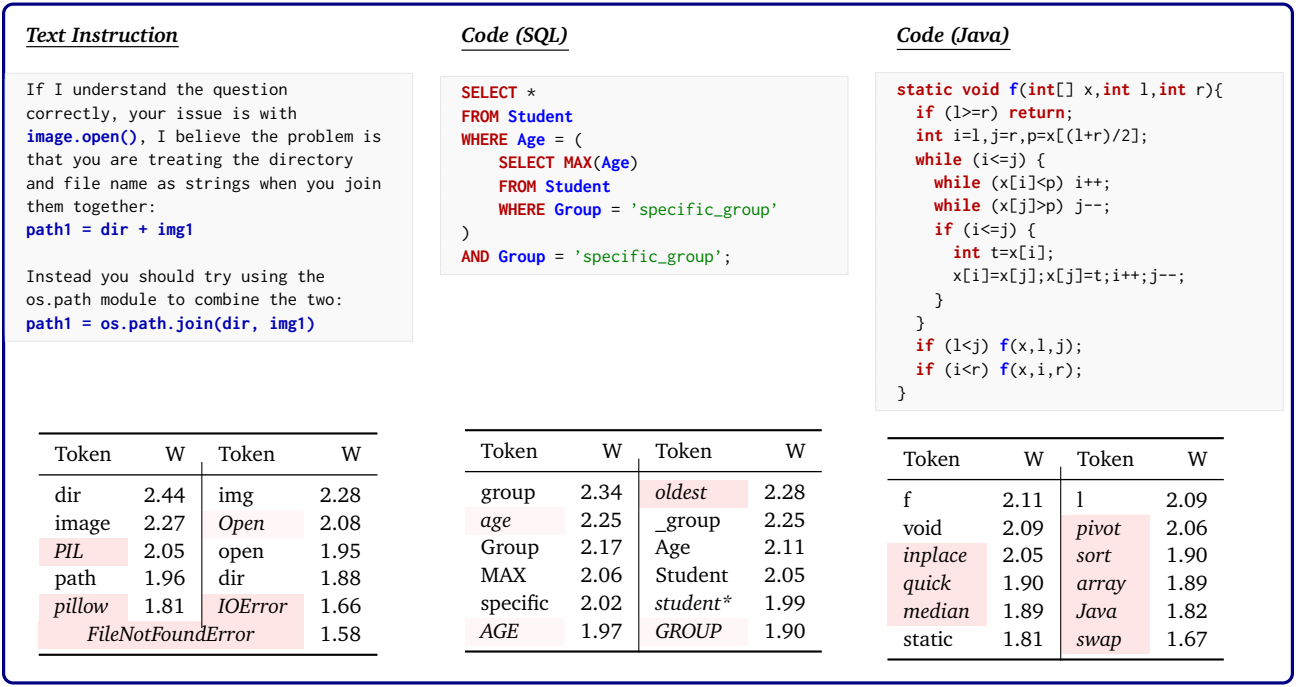

\centering

\begin{tcolorbox}[
    colframe=blue!50!black,
    colback=white,
    boxrule=1.2pt,
    arc=3pt,
    width=\textwidth,
    enhanced,
    left=5pt,
    right=5pt,
    top=5pt,
    bottom=5pt,
]

\begin{minipage}[t]{0.3\linewidth}
\footnotesize
\textbf{\textit{\underline{Text Instruction}}}
\vspace{3mm}

\begin{lstlisting}[style=instructionstyle]
If I understand the question correctly, your issue is with (*@\textcolor{blue!70!black}{\bfseries image.open()}@*), I believe the problem is that you are treating the directory and file name as strings when you join them together:
(*@\textcolor{blue!70!black}{\bfseries path1 = dir + img1}@*)

Instead you should try using the os.path module to combine the two:
(*@\textcolor{blue!70!black}{\bfseries path1 = os.path.join(dir, img1)}@*)
\end{lstlisting}

\vspace{11mm}
\centering
\hspace*{-2mm}
\begin{tabular}{l r | l r}
\toprule
Token & W & Token & W \\
\midrule
dir        & 2.44 & img        & 2.28 \\
image      & 2.27 & \cellcolor{red!3}\textit{Open}       & 2.08 \\
\cellcolor{red!10}\textit{PIL}        & 2.05 & open       & 1.95 \\
path       & 1.96 & dir        & 1.88 \\
\cellcolor{red!10}\textit{pillow}     & 1.81 & \cellcolor{red!10}\textit{IOError}    & 1.66 \\
\multicolumn{3}{c}{\cellcolor{red!10}\textit{FileNotFoundError}}         &   1.58   \\
\bottomrule
\end{tabular}
\end{minipage}
\hfill
\begin{minipage}[t]{0.3\linewidth}
\footnotesize
\textbf{\textit{\underline{Code (SQL)}}}
\vspace{2mm}

\begin{lstlisting}[style=sqlstyle]
SELECT *
FROM Student
WHERE Age = (
    SELECT MAX(Age)
    FROM Student
    WHERE Group = 'specific_group'
)
AND Group = 'specific_group';
\end{lstlisting}

\vspace{18mm}
\centering
\hspace*{-2mm}
\begin{tabular}{l r | l r}
\toprule
Token & W & Token & W \\
\midrule
group      & 2.34 & \cellcolor{red!10}\textit{oldest}     & 2.28 \\
\cellcolor{red!3}\textit{age}     & 2.25 & \_group         & 2.25 \\
Group      & 2.17 & Age         & 2.11 \\
MAX        & 2.06 & Student     & 2.05 \\
specific   & 2.02 & \textit{student*}     & 1.99 \\
\cellcolor{red!3}\textit{AGE}        & 1.97 & \cellcolor{red!3}\textit{GROUP}       & 1.90 \\
\bottomrule
\end{tabular}
\end{minipage}
\hfill
\begin{minipage}[t]{0.3\linewidth}
\footnotesize
\textbf{\textit{\underline{Code (Java)}}}
\vspace{1.5mm}

\begin{lstlisting}[style=javastyle]
static void f(int[] x,int l,int r){
  if (l>=r) return;
  int i=l,j=r,p=x[(l+r)/2];
  while (i<=j) {
    while (x[i]<p) i++;
    while (x[j]>p) j--;
    if (i<=j) {
      int t=x[i]; 
      x[i]=x[j];x[j]=t;i++;j--;
    }
  }
  if (l<j) f(x,l,j);
  if (i<r) f(x,i,r);
}
\end{lstlisting}

\vspace{1mm}
\hspace*{-2mm}
\begin{tabular}{l r | l r}
\toprule
Token & W & Token & W \\
\midrule
f        & 2.11 & l   & 2.09 \\
void     & 2.09 & \cellcolor{red!10}\textit{pivot}        & 2.06 \\
\cellcolor{red!10}\textit{inplace}   & 2.05 & \cellcolor{red!10}\textit{sort}    & 1.90 \\
\cellcolor{red!10}\textit{quick}  & 1.90 & \cellcolor{red!10}\textit{array}    & 1.89 \\
\cellcolor{red!10}\textit{median}    & 1.89 & \cellcolor{red!10}\textit{Java}   & 1.82 \\
static         & 1.81 &   \cellcolor{red!10}\textit{swap}     & 1.67 \\
\bottomrule
\end{tabular}
\end{minipage}

\end{tcolorbox}

\caption{
Examples of sparse vocabulary activations produced by SPLADE-Code-8B for different inputs. 
Highest-weighted \colorbox{red!10}{expansion tokens} in the representations capturing query intent, or algorithmic concepts highlighted.
}
\label{fig:examples}
\end{figure*}

%% file: tables/backbone.tex
\begin{table}[t]
\centering
\adjustbox{max width=0.48\textwidth}{
\begin{tabular}{lcccc}
\toprule
\multirow{2}{*}{Model Variant}  &\multirow{2}{*}{CoIR} & MTEB & Code & \multirow{2}{*}{CPRet}  \\
      &    &   Code &   RAG  &      \\
\midrule

\multicolumn{5}{l}{\textit{\textbf{SPLADE-Code w/ Generalist LLM}}} \\[0.3em]
\hspace{1em} Qwen3-8B & \textbf{76.7} & \textbf{77.8} & {68.5} & \textbf{76.8} \\
\hspace{1em} Llama3-8B & \underline{76.6} & \underline{77.6} & \underline{68.6} & 72.4 \\

\midrule
\multicolumn{5}{l}{\textit{\textbf{SPLADE-Code w/ Code-Specialized LLM}}} \\[0.3em]
\hspace{1em} Qwen2.5Coder-0.5B & 70.4 & 71.5 & 62.4 & 46.6 \\
\hspace{1em} Qwen2.5Coder-1.5B & 71.9 & 72.9 & 63.9 & 50.4 \\
\hspace{1em} Qwen2.5Coder-7B & 76.4 & \underline{77.6} & \textbf{71.3} & 74.9 \\

\midrule
\multicolumn{5}{l}{\textit{\textbf{SPLARE-Code w/ Sparse Auto-Encoder (\citeauthor{formal2026learning})}}} \\[0.3em]
\hspace{1em} with general-domain SAE & 76.0 & 77.1 & 62.3 & 67.8 \\
\hspace{1em} with code-specific SAE & 75.2 & 76.4 & 69.4 & 72.2 \\

\bottomrule
\end{tabular}
}
\caption{Ablation comparison of LSR models with different LLM backbones. Results are reported without checkpoint merging for fair comparison.}
\label{tab:backbone}
\vspace{0.5em}
\end{table}

%% file: tables/ablations.tex
\begin{table}[t]
\centering
\small
\adjustbox{max width=0.48\textwidth}{
\begin{tabular}{lcc}
\toprule
Model Variant & CoIR & MTEB Code \\
\midrule
\textit{\textbf{SPLADE-Code-0.6B}} & & \\
\hspace{1em} (Base) & 72.6 & 73.5 \\
\hspace{1em} (w/ Instructions) & 73.0 & 73.4 \\
\hspace{1em} (English$\rightarrow$Code) & 67.1 & 68.5 \\
\hspace{1em} (Contrastive) & 68.3 & 67.4 \\
\bottomrule
\end{tabular}
}
\caption{Ablations on SPLADE-Code-0.6B. The default setup uses KL distillation without instructions or intermediate English fine-tuning. Results are reported without checkpoint merging for fair comparison.}
\label{tab:lsr_ablation_code}
\end{table}

%% file: paper/6_conclusion.tex
\section{Conclusion}
\label{sec:conclusion}

We introduced SPLADE-Code, the first learned sparse retrieval models specialized for code. Across multiple in-domain and out-of-domain benchmarks, SPLADE-Code achieves strong effectiveness with a lightweight single-stage training pipeline, showing that LSR is a viable alternative to dense retrieval for code search. Our analysis further shows that expansion terms are central to this performance, enabling sparse representations to bridge lexical matching and semantic abstraction while remaining interpretable. We also showed that SPLADE-Code can provide favorable effectiveness--efficiency trade-offs, reaching below 1\,ms latency with only limited effectiveness loss, and is interpretable.

Future work could integrate LSR more tightly into real-world agentic software engineering systems, where retrieval must jointly satisfy effectiveness, latency, and interpretability constraints.





\vspace{1em}

\section{Limitations}
\label{sec:limitations}

Our study has several limitations. First, while we evaluate SPLADE-Code across multiple benchmarks and both in-domain and out-of-domain settings, these datasets may not fully reflect the distribution of real developer queries (e.g., repository-specific naming conventions, evolving APIs, or interactive debugging contexts). Additional evaluation on more complex and realistic datasets, query logs and repository-level tasks would strengthen external validity.
Then, we primarily study retriever-level effectiveness. In practical LLM-based software engineering systems, retrieval interacts with reranking, tool use, and generation. Measuring end-to-end impacts (e.g., downstream task success, faithfulness, and debugging outcomes) remains an important direction. Finally, a further challenge in comparing latency is that dense retrieval is often optimized for multi-CPU or GPU-based similarity search, while sparse inverted-index methods are typically evaluated as single-core systems designed to handle large query volumes in parallel, so the resulting measurements are not directly equivalent.

%% file: supplementary/A_experiments.tex
\newpage 

\appendix

\section{Hyper-parameters}
\label{sec:appendix:hyperparam}

\header{Training settings.} Table~\ref{tab:hyper} summarizes the main hyper-parameters we used in our experiments. We train our model on the training set of CoIR, containing of ten datasets merged together for a total of 2.2M training samples, for one epoch. Each training sample consists of one query, one positive, and $n$ negatives. In practice, we use batch size of 256, achieved with gradient accumulation, and 7 or 8 negatives per query depending on the parameter size of the models. We used LoRA~\cite{hu2022lora} rank of 64.
The negatives are mined from retrieval made with Qwen3 Embedding (0.6B), and scored with  Qwen3 Reranker (4B) to be then distilled~\cite{zhang2025qwen3}. We select the negative in-between the rank 50 to 100 following previous training of embedding models. All models are trained with a max length of 512. Since queries and documents are very long, we observed that models tend to produce high similarity scores, we thus use a high temperature in the KLD loss to mitigate and have more stable training, we used temperature of 300. The learning rate is set to 1e-4.
Model merging is done with weighted spherical merging~\cite{yang2024modelmerging} from three checkpoints: (1) the same base model after the first epoch, (2) and after a second epoch of training, (3) the base model trained with a max context length of 1024 for one epoch. Small models (0.6B) also contain a fourth checkpoint trained with full finetuning (no LoRA).
All decoder models are pretrained on MS MARCO passages~\cite{msmarco} with bi-directional attention to transform from causal to bi-directional attention models.

\header{Dense models training settings.} In \S \ref{sec:exps}, we train dense models in controlled experiments, to enable direct comparison with LSR models. These models are trained using \texttt{eos} token pooling. The learning rate is $1e-5$, the batch size is $128$, the warmup ratio is $0.05$ and the KL temperature is set to 0.1 (after some hyper-parameter search). 8 negatives are used for each query. The rest of the parameters are identical to the sparse models.


\input{tables/hyperparam}

\header{Masked Language pretraining} SPLADE models use bi-directional attention. The models are first fine-tuned quickly with bi-directional attention and a masked language modelling objective. We use the exact same setup as in \cite{formal2026learning,zeng2025scaling}. Note that this step is not very expensive: it requires about 15 hours on a single GPU for the 8B model.

\header{Hardware.} Training is done on 4 A100 with 80GB of memory each. Retrieval latencies are computed on a single AMD EPYC 7313 processor for fair comparison between dense and LSR.

\header{Latency experiments.} For the LSR retrieval we used the \textit{seismic} library, which contains an efficient implementation of inverted index~\cite{bruch2024efficient}. In particular, for the latency experiments, we use \textit{seismic} with parameters $k=1000$, $\mathit{query\_cut}=500$, $\mathit{heap\_factor}=2.5$, and $\mathit{n\_knn}=0$. For $hnsw$, we use faiss \cite{douze2025faiss} and parameters $\mathit{M}=32$, $\mathit{efConstruction}=40$ and vary $\mathit{efSearch}$. We retrieve $\mathit{top\_k}=1000$ in all cases.

\section{Datasets Statistics}
\label{sec:appendix:stat}

We provide additional statistics about the benchmark in Table~\ref{table_stat_bench}, including the number of datasets for each benchmark, the list of programming languages and the main tasks within each benchmark. We also provide some dataset sizes for queries and collections.
\input{tables/stat}

\section{Additional Analysis}

\begin{figure}[t]
    \centering
    \includegraphics[width=\linewidth]{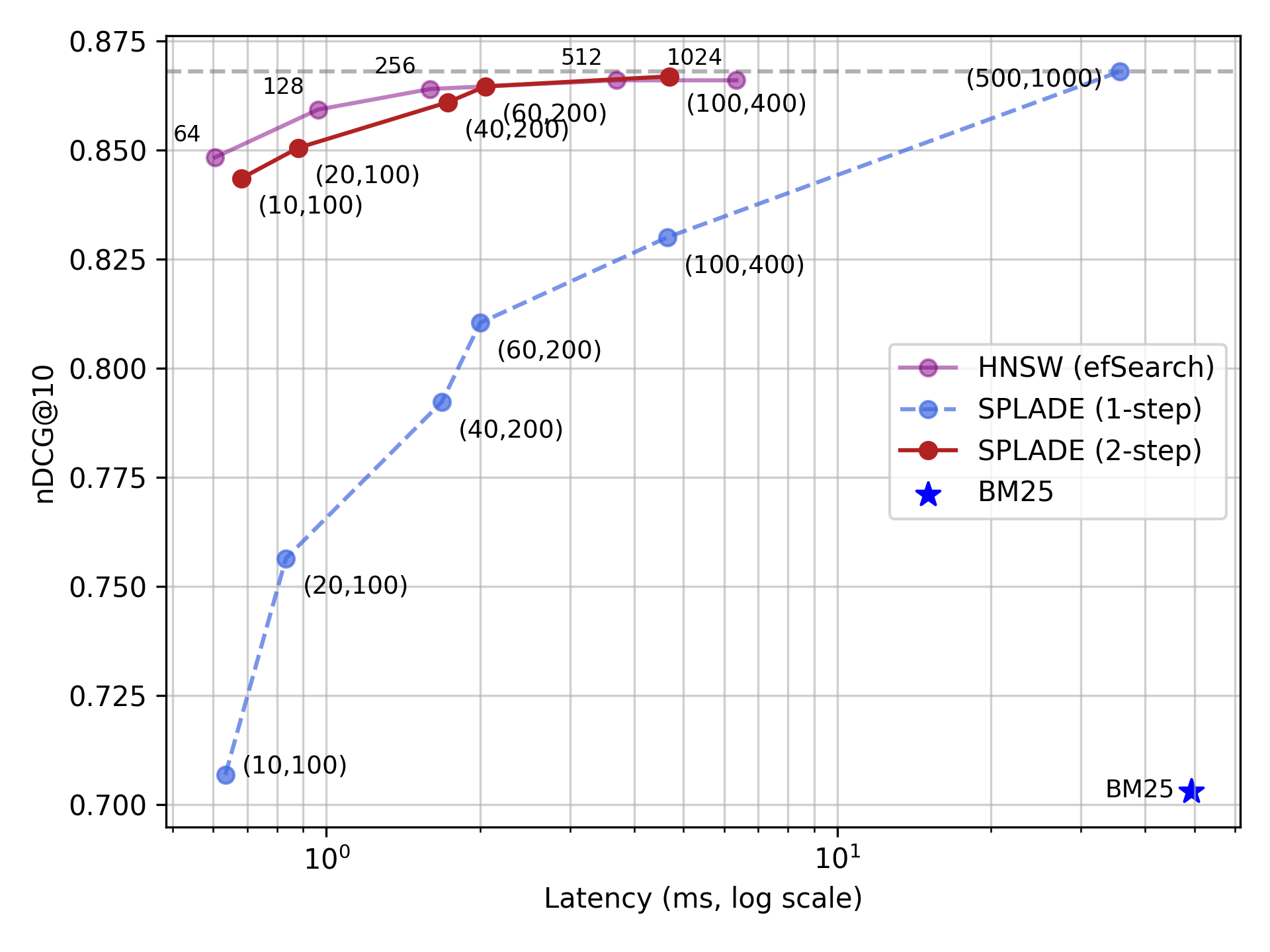}
    \caption{Latency of SPLADE-Code-0.6B for code representation with two-step SPLADE on Qwen3-0.6B. On Qwen3-0.6B the dot product for the dense is more efficient since embedding dimensions are lower, hence lower latencies with HNSW, but SPLADE still matches the Pareto frontier.}
    \label{fig:lsrcode06b}
\end{figure}

\header{Efficiency.} We provide in Figure~\ref{fig:lsrcode06b} an additional comparison of the latency of SPLADE-Code-0.6B. Latencies are overall similar to the curve provided in the main body of the paper for SPLADE-Code-8B. Also note that we measure here the latency of the retrieval only from passing through the inverted index or the vector search algorithm, we do not consider the latency from encoding the query and getting the sparse representation, since this time is similar for dense and LSR models. Also note that the added latency from two-step SPLADE is estimated here as the time to compute 1000 dot products, which we estimated to 50 microseconds for 1000 dot products between vectors of size 100K with only 400 non zero dimension.

\header{Ablations.}
We provide in Table~\ref{tab:codemteb_app} and Table~\ref{tab:coderag_cpret_merged_app} the performance of some models on individual datasets from each benchmark. This includes some models from ablation or models without checkpoint merging.

\input{tables/full_indomain_app}

\input{tables/full_ood_app}

\section{Performance on English and Multilingual Benchmark}

While the main focus of our research was to develop code retrieval model, we observed that the models were good in English and had some multilingual capacities. We provide in Table~\ref{tab:mteb_eng} and Table~\ref{tab:mteb_multi} the performance of SPLADE-Code on MTEB English and MTEB multilingual. As a comparison, LION-SP-8B~\cite{zeng2025scaling} achieves 48.5 on MTEB(eng, v2) and 50.0 on MTEB(Multilingual, v2). SPLADE-Code-8B is more effective than strong monolingual English retrieval modes. SPLADE-EN-0.6B is the English SPLADE model we trained for the intermediate English finetuning ablation within Table~\ref{tab:lsr_ablation_code}. It achieves good performance in English.

\input{tables/mteb_eng_multilingual_app}

%% file: tables/hyperparam.tex
\begin{table*}[t]
\centering
\small
\begin{tabular}{lccc}
\toprule
 & \textbf{SPLADE-Code 0.6B} & \textbf{SPLADE-Code 1.7B} & \textbf{SPLADE-Code 8B} \\
\midrule
Base Model & Qwen/Qwen3-0.6B & Qwen/Qwen3-1.7B & Qwen/Qwen3-8B \\
Batch size & 256 & 256 & 256 \\
Negatives per query & 8 & 7 & 7 \\
LoRA Rank & 64 & 64 & 64 \\
Max Length & 512 & 512 & 512 \\
Learning Rate & $1\times 10^{-4}$ & $1\times 10^{-4}$ & $1\times 10^{-4}$ \\
Temperature (KLD) & 300 & 300 & 300 \\
Epochs & 1 & 1 & 1 \\
Bidirectional Attention & Yes & Yes & Yes \\
Warmup Ratio & 0.01 & 0.01 & 0.01 \\
Quantization & BF16 & BF16 & BF16 \\
Lambda q & $1\times 10^{-4}$ & $1\times 10^{-4}$ & $1\times 10^{-4}$ \\
Lambda d & $1\times 10^{-4}$ & $1\times 10^{-4}$ & $1\times 10^{-4}$ \\
KLD Loss Weight & 0.5 & 0.5 & 0.5 \\
Weight Decay & 0.0 & 0.0 & 0.0 \\
Optimizer & Adam & Adam & Adam \\
\midrule
Inference Max Length & 2048 & 2048 & 2048 \\
Pruning Inference & (500,1000) & (500,1000) & (500,1000) \\
Heap Factor (LSR) & 2.5 & 2.5 & 2.5 \\
\bottomrule
\end{tabular}
\vspace{1mm}
\caption{Main training configurations for the base models. For model merging we merge the same model after 1 and 2 epochs, and with a max length of 1024 (3 ckpts). For 0.6B we also merge a full finetuning model.}
\label{tab:hyper}
\end{table*}

%% file: tables/stat.tex
\begin{table*}[t]
\centering
\footnotesize
\setlength{\tabcolsep}{4pt}
\renewcommand{\arraystretch}{1.2}
\adjustbox{max width=\textwidth}{
\begin{tabular}{p{5.0cm} c c p{3.0cm} p{3.0cm} c c}
\toprule
\textbf{Benchmark} 
& \#Datasets
& \#Lang
& \textbf{Task / Domain} 
& \textbf{Languages} 
& \textbf{Test Size} 
& \textbf{Collection Size} \\
\midrule

\textbf{CoIR} \cite{li-etal-2025-coir}
& 10
& 14
& Code Retrieval \newline
  Text-to-Code \newline
  Code-to-Code \newline
  Code-to-Text \newline
  Hybrid Code \newline
  (8 Coding Tasks)
& Python, SQL, Go, Java, JS, C, CSS, PHP, C++, HTML, Rust, Shell, Swift, Ruby
& 180 -- 53K 
&  816 -- 1M \\
\midrule

\textbf{CodeRAG-Bench} \cite{wang2025coderag}
& 6
& 1
& Retrieval-Augmented Code Generation \newline
  Open-Domain  \newline
  Basic Programming \newline
  Repository-Level \newline
  Retrieval \newline
  (8 Coding Tasks)
& Python
& 164 -- 1K 
& 237 -- 40.9K  \\
\midrule

\textbf{CPRet} \cite{deng2025cpret}
& 4
& 20+
& Competitive Programming Retrieval \newline
  Text-to-Code \newline
  Code-to-Code \newline
  Problem-to-Duplicate \newline
  Simplified-to-Full \newline
  (4 Coding Tasks)
& Python, C++, C, Java, Perl, Groovy, R, Rust, Dart, C\#, Matlab, \newline
  CoffeeScript, Assembly, Haskell, Batchfile, Go, Ruby, JavaScript, Kotlin
& 168 -- 10K 
&  10K -- 41.6K \\
\midrule
\textbf{MTEB-Code} (build upon CoIR) \newline
\cite{muennighoff-etal-2023-mteb} & 12 & 14 & Code Retrieval \newline
  Text-to-Code \newline
  Code-to-Code \newline
  Code-to-Text \newline
  Hybrid Code \newline
  (9 Coding Tasks)
& Python, SQL, Go, Java, JS, C, CSS, PHP, C++, HTML, Rust, Shell, Swift, Ruby
& 180 -- 53K 
&  816 -- 1M \\
\bottomrule
\end{tabular}
}
\caption{Benchmark statistics of CoIR, CodeRAG-Bench, CPRet and MTEB Code. 
Test and collection sizes are reported as ranges (min--max per dataset). MTEB Code is composed of CoIR with the two additional datasets: CodeEditRetrieval and a variation of CodeSearchNet.}
\label{table_stat_bench}
\end{table*}

%% file: tables/full_indomain_app.tex
\begin{table*}[t!]
\centering
\adjustbox{max width=\textwidth}{
\begin{tabular}{lcccccccccccc|cc}
\toprule
\multirow{2}{*}{Model} &
\multirow{2}{*}{Apps} &
\multirow{2}{*}{CosQA} &
\multirow{2}{*}{T2SQL} &
\multirow{2}{*}{CSN$^\dagger$} &
\multirow{1}{*}{CSN} &
\multirow{1}{*}{CSN} &
\multirow{1}{*}{CodeEdit} &
\multicolumn{2}{c}{CodeTrans} &
\multirow{1}{*}{StackOv} &
\multicolumn{2}{c}{CodeFeedback} &
\multirow{1}{*}{\textbf{Avg}} &
\multirow{1}{*}{\textbf{Avg}} \\
&
& & & & -COIR & -CCR &
Retrieval$^\dagger$ &
-Contest &
-DL &
QA &
-ST &
-MT &
\textbf{COIR} &
\textbf{MTEB}
\\
\midrule


BM25 & 4.7 & 18.7 & 24.9 & 63.5 & 70.3 & 59.3 & 53.3  & 47.7 & {34.4} & 70.2 & 68.1 & 59.1 & \cellcolor{gray!10}45.8 & \cellcolor{gray!10}47.9 \\

\midrule

SPLADE-lex.-0.6B
& 10.2 & 6.8 & 18.7 & 42.7 & 37.5 & 45.1 & 35.0
& 55.5 & 30.9 & 71.1 & 67.7 & 79.9
& \cellcolor{gray!10}41.8 & \cellcolor{gray!10}42.3 \\

SPLADE-lex.-1.7B
& 12.8 & 7.7 & 17.9 & 46.3 & 39.0 & 48.9 & 35.9
& 55.2 & 29.7 & 73.7 & 68.1 & 82.0
& \cellcolor{gray!10}43.1 & \cellcolor{gray!10}43.5 \\

SPLADE-lex.-8B
& 16.8 & 7.4 & 18.6 & 49.7 & 41.9 & 55.3 & 37.6
& 61.5 & 29.2 & 76.5 & 69.5 & 85.3
& \cellcolor{gray!10}45.8 & \cellcolor{gray!10}46.2 \\

\midrule

SPLADE-Code-no-merging-0.6B
& 67.3 & 33.4 & 69.0 & 90.3 & 87.6 & 94.6 & 65.5
& 89.2 & 23.7 & 93.4 & 77.1 & 90.3
& \cellcolor{gray!10}72.6  & \cellcolor{gray!10}73.4 \\

SPLADE-Code-no-merging-1.7B
& 72.8 & 33.8 & 71.7 & 91.1 & 88.1 & 94.2 & 67.4
& 90.9 & 25.6 & 94.0 & 80.5 & 90.6
& \cellcolor{gray!10}74.2 & \cellcolor{gray!10}75.1  \\

SPLADE-Code-no-merging-8B
& 87.8 & 34.0 & 71.1 & 92.1 & 89.9 & 96.7 & 74.8
& 93.9 & 23.1 & 96.4 & 81.1 & 91.9
& \cellcolor{gray!10}76.6 & \cellcolor{gray!10}77.7  \\

\midrule

SPLADE-Code-QwenCoder-0.5B
& 50.5 & 32.8 & 70.3 & 90.0 & 87.2 & 93.5 & 63.2
& 86.7 & 24.6 & 91.8 & 79.8 & 87.3
& \cellcolor{gray!10}70.4 & \cellcolor{gray!10}71.5  \\

SPLADE-Code-QwenCoder-1.5B
& 54.3 & 30.5 & 72.7 & 90.6 & 87.6 & 92.2 & 64.9
& 87.7 & 31.4 & 92.6 & 81.7 & 88.6
& \cellcolor{gray!10}71.9 & \cellcolor{gray!10}72.9  \\

SPLADE-Code-QwenCoder-7B
& 86.2 & 32.3 & 70.6 & 92.3 & 90.2 & 96.9 & 74.4
& 94.4 & 23.5 & 96.1 & 81.5 & 92.2
& \cellcolor{gray!10}76.4 & \cellcolor{gray!10}77.6  \\

\midrule

SPLARE-LlamaScope-7B-L26
& 78.0 & 32.6 & 75.0 & 90.2 & 88.2 & 95.1 & 75.1
& 93.3 & 27.2 & 95.7 & 83.9 & 90.6 & \cellcolor{gray!10}76.0 & \cellcolor{gray!10}77.1 \\

SPLARE-Llama3-SAE-CoIR-7B-L24
& 81.0 & 32.3 & 70.6 & 91.6 & 89.0 & 95.7 & 74.1
& 93.9 & 20.9 & 96.7 & 74.6 & 91.7 & \cellcolor{gray!10}74.6 & \cellcolor{gray!10}76.0 \\

SPLARE-Qwen3-SAE-CoIR-8B-L32
& 87.6 & 28.2 & 70.7 & 90.8 & 89.1 &96.0 & 73.5
& 93.1 & 22.9 & 96.2 & 77.2 & 91.4
& \cellcolor{gray!10}75.2 & \cellcolor{gray!10}76.4 \\


\midrule

SPLADE-Code-0.6B
& 66.5 & 35.4 & 74.3 & 90.5 & 86.2 & 91.4 & 68.4
& 90.9 & 31.4 & 93.3 & 84.7 & 92.0
 & \cellcolor{gray!10}74.6 & \cellcolor{gray!10}75.4 \\

SPLADE-Code-1.7B
& 71.1 & 32.1 & 73.3 & 91.3 & 87.9 & 92.8 & 66.7
& 92.3 & 25.4 & 93.9 & 84.1 & 91.4
& \cellcolor{gray!10}74.4 & \cellcolor{gray!10}75.2  \\

SPLADE-Code-8B
& 86.7 & 32.5 & 75.7 & 92.1 & 88.9 & 94.6 & 77.1
& 94.6 & 28.1 & 96.5 & 87.6 & 93.9
& \cellcolor{gray!10}77.9  & \cellcolor{gray!10}79.0 \\

\bottomrule
\end{tabular}
}
\caption{Detailed MTEB Code retrieval results of ablations (nDCG@10), using pruning $(500,1000)$.}
\label{tab:codemteb_app}
\end{table*}

%% file: tables/full_ood_app.tex
\begin{table*}[t]
\centering
\adjustbox{max width=\textwidth}{
\begin{tabular}{lccccccc|ccccc}
\toprule

\multirow{2}{*}{Model} 
& \multicolumn{7}{c}{\textbf{CodeRAG}} 
& \multicolumn{5}{c}{\textbf{CPRet}} \\

\cmidrule(r){2-8} \cmidrule(l){9-13}

& Human & \multirow{2}{*}{MBPP} & \multirow{2}{*}{DS-1000} & \multirow{2}{*}{ODEX} 
& Repo & SWE & \multirow{2}{*}{\textbf{Avg}}
& \multirow{2}{*}{T2C} & \multirow{2}{*}{C2C} 
& \multirow{2}{*}{P2Dup} & \multirow{2}{*}{S2Full} 
& \multirow{2}{*}{\textbf{Avg}} \\

& Eval &  &  &  & Eval & Bench-L &  &  &  &  &  &  \\

\midrule

BM25 
& 100.0 & 98.6 & 5.2 & 6.7 & 93.2 & 43.0 & \cellcolor{gray!10}57.7
& 0.9 & 7.4 & 12.4 & 53.8 & \cellcolor{gray!10}18.6 \\

\midrule

SPLADE-lex.-0.6B
& 100.0 & 99.0 & 5.2 & 1.7 & 94.1 & 33.5 & \cellcolor{gray!10}55.6
& 7.1 & 13.0 & 20.1 & 58.1 & \cellcolor{gray!10}24.6 \\

SPLADE-lex.-1.7B
& 99.8 & 99.2 & 4.8 & 2.9 & 94.3 & 28.0 & \cellcolor{gray!10}54.8
& 8.9 & 16.2 & 22.8 & 59.8 & \cellcolor{gray!10}26.9 \\

SPLADE-lex.-8B
& 99.8 & 98.4 & 6.4 & 3.2 & 93.9 & 38.9 & \cellcolor{gray!10}56.7
& 11.6 & 19.5 & 23.1 & 63.4 & \cellcolor{gray!10}29.4 \\

\midrule

SPLADE-Code-no-merging-0.6B
& 100.0 & 98.9 & 22.8 & 27.0 & 95.1 & 38.2 & \cellcolor{gray!10}63.7
& 40.4 & 50.2 & 43.8 & 91.0 & \cellcolor{gray!10}56.4 \\

SPLADE-Code-no-merging-1.7B
& 100.0 & 98.9 & 29.0 & 28.3 & 93.4 & 39.9 & \cellcolor{gray!10}64.9
& 46.7 & 57.6 & 52.9 & 92.8 & \cellcolor{gray!10}62.5 \\

SPLADE-Code-no-merging-8B 
& 100.0 & 98.6 & 38.1 & 36.7 & 90.8 & 47.0 & \cellcolor{gray!10}68.5
& 70.7 & 76.8 & 64.1 & 95.8 & \cellcolor{gray!10}76.8 \\

\midrule

SPLADE-Code-QwenCoder-0.5B
& 100.0 & 99.1 & 24.0 & 21.0 & 93.3 & 36.8 & \cellcolor{gray!10}62.4
& 28.1 & 35.5 & 37.2 & 85.4 & \cellcolor{gray!10}46.6 \\

SPLADE-Code-QwenCoder-1.5B
& 100.0 & 99.3 & 24.0 & 22.4 & 93.2 & 44.9 & \cellcolor{gray!10}63.9
& 30.0 & 41.6 & 42.3 & 87.7 & \cellcolor{gray!10}50.4 \\

SPLADE-Code-QwenCoder-7B
& 100.0 & 98.6 & 36.5 & 36.0 & 90.2 & 66.7 & \cellcolor{gray!10}71.3
& 70.5 & 73.4 & 59.9 & 95.6 & \cellcolor{gray!10}74.9 \\

\midrule

SPLARE-LlamaScope-7B-L26
& 100.0 & 93.4 & 30.2 & 24.2 & 75.0 & 50.8 & \cellcolor{gray!10}62.3
& 59.4 & 66.6 & 51.7 & 93.5 & \cellcolor{gray!10}67.8 \\

SPLARE-Llama3-SAE-CoIR-7B-L24
& 100.0 & 98.3 & 37.4 & 38.5 & 92.9 & 49.3 & \cellcolor{gray!10}69.4
& 63.9 & 72.2 & 58.1 & 94.4 & \cellcolor{gray!10}72.2 \\

\midrule

SPLADE-Code-0.6B
& 100.0 & 99.2 & 21.1 & 29.1 & 95.8 & 39.5 & \cellcolor{gray!10}64.1
& 42.2 & 47.0 & 50.2 & 92.7 & \cellcolor{gray!10}58.0 \\

SPLADE-Code-1.7B
& 100.0 & 99.0 & 30.7 & 28.4 & 93.0 & 42.5 & \cellcolor{gray!10}65.6
& 48.1 & 56.8 & 55.4 & 93.5 & \cellcolor{gray!10}63.4 \\

SPLADE-Code-8B
& 100.0 & 99.4 & 42.0 & 38.8 & 94.2 & 52.6 & \cellcolor{gray!10}71.2
& 70.9 & 76.4 & 66.3 & 96.1 & \cellcolor{gray!10}77.4 \\

\bottomrule
\end{tabular}
}
\caption{Detailed out-of-domain retrieval performance on CodeRAG and CPRet for ablation models (nDCG@10). Retrieval with pruning $(1000,1000)$.}
\label{tab:coderag_cpret_merged_app}
\end{table*}

%% file: tables/mteb_eng_multilingual_app.tex
\begin{table*}[t]
\centering
\footnotesize
\setlength{\tabcolsep}{4pt}
\begin{tabular}{lcc|c}
\toprule
\textbf{Dataset} 
& \textbf{SPLADE-Code-0.6B} 
& \textbf{SPLADE-Code-8B} 
& \textbf{SPLADE-EN-0.6B} \\
\midrule
\textbf{Average}                     & \cellcolor{gray!10}45.4 & \cellcolor{gray!10}49.7 & \cellcolor{gray!10}48.1 \\
\midrule
ArguAna                              & 56.4 & 59.9 & 52.3 \\
CQADupstackGamingRetrieval           & 56.8 & 59.2 & 56.0 \\
CQADupstackUnixRetrieval             & 40.5 & 46.7 & 37.6 \\
ClimateFEVERHardNegatives            & 19.0 & 21.7 & 19.6 \\
FEVERHardNegatives                   & 60.8 & 66.0 & 62.2 \\
FiQA2018                             & 36.9 & 47.8 & 36.0 \\
HotpotQAHardNegatives                & 47.8 & 52.9 & 59.2 \\
SCIDOCS                              & 17.4 & 19.7 & 16.6 \\
TRECCOVID                            & 68.6 & 69.1 & 77.3 \\
Touche2020Retrieval.v3               & 49.8 & 53.5 & 63.9 \\
\bottomrule
\end{tabular}
\caption{Results of the SPLADE-Code models on  MTEB(eng, v2) with pruning $(40,400)$.}
\label{tab:mteb_eng}
\end{table*}

\begin{table*}[t]
\centering
\footnotesize
\setlength{\tabcolsep}{4pt}
\begin{tabular}{lcc|c}
\toprule
\textbf{Dataset} 
& \textbf{SPLADE-Code-0.6B} 
& \textbf{SPLADE-Code-8B} 
& \textbf{SPLADE-EN-0.6B} \\
\midrule
\textbf{Average}                              & \cellcolor{gray!10}48.3 & \cellcolor{gray!10}54.3 & \cellcolor{gray!10}47.0 \\
\midrule
AILAStatutes                         & 28.1 & 32.4 & 27.7 \\
ArguAna                              & 56.4 & 59.9 & 52.3 \\
BelebeleRetrieval                    & 54.1 & 77.1 & 48.4 \\
CovidRetrieval                       & 73.3 & 72.7 & 78.9 \\
HagridRetrieval                      & 98.7 & 98.7 & 98.8 \\
LEMBPasskeyRetrieval                 & 38.8 & 38.8 & 38.8 \\
LegalBenchCorporateLobbying          & 94.1 & 95.7 & 93.1 \\
MIRACLRetrievalHardNegatives         & 31.5 & 43.2 & 31.5 \\
MLQARetrieval                        & 55.7 & 71.5 & 56.6 \\
SCIDOCS                              & 17.4 & 19.7 & 16.6 \\
SpartQA                              & 4.5  & 2.9  & 1.1 \\
StackOverflowQA                      & 86.9 & 93.5 & 74.7 \\
StatcanDialogueDatasetRetrieval      & 28.0 & 37.2 & 22.3 \\
TRECCOVID                            & 68.6 & 69.1 & 77.3 \\
TempReasonL1                         & 0.2  & 0.2  & 0.3 \\
TwitterHjerneRetrieval               & 49.7 & 71.9 & 42.5 \\
WikipediaRetrievalMultilingual       & 80.8 & 89.1 & 80.8 \\
WinoGrande                           & 2.9  & 4.8  & 3.7 \\
\bottomrule
\end{tabular}
\caption{Results of the SPLADE-Code models on  MTEB(Multilingual, v2) with pruning $(40,400)$.}
\label{tab:mteb_multi}
\end{table*}